\documentclass[
aps,
prd,
showpacs,
preprint,
tightenlines,
superscriptaddress,
nofootinbib,
floatfix]
{revtex4}
\usepackage{graphicx,
longtable
}

\begin{document}
%
\newcommand{\Lesssim}{\stackrel{\textstyle <}{\raisebox{-.6ex}{$\sim$}}}
\newcommand{\Gtrsim}{\stackrel{\textstyle >}{\raisebox{-.6ex}{$\sim$}}}

\title{    Analysis of energy- and time-dependence\\
  of supernova shock effects on neutrino crossing probabilities}
%
\author{        G.L.~Fogli}
\affiliation{   Dipartimento di Fisica
                and Sezione INFN di Bari\\
                Via Amendola 173, 70126 Bari, Italy\\}
\author{        E.~Lisi}
\affiliation{   Dipartimento di Fisica
                and Sezione INFN di Bari\\
                Via Amendola 173, 70126 Bari, Italy\\}
\author{        A.~Mirizzi}
\affiliation{   Dipartimento di Fisica
                and Sezione INFN di Bari\\
                Via Amendola 173, 70126 Bari, Italy\\}
\author{        D.~Montanino}
\affiliation{   Dipartimento di Scienza dei Materiali
                and Sezione INFN di Lecce\\
                Via Arnesano, 73100 Lecce,    Italy\\}

\begin{abstract}
It has recently been realized that supernova neutrino signals may
be affected by shock propagation over a time interval of a few
seconds after bounce. In the standard three-neutrino oscillation
scenario, such effects crucially depend on the neutrino level
crossing probability $P_H$ in the 1-3 sector. By using a
simplified parametrization of the time-dependent supernova radial
density profile, we explicitly show that simple analytical
expressions for $P_H$ accurately reproduce the phase-averaged
results of numerical calculations in the relevant parameter space.
Such expressions are then used to study the structure of $P_H$ as
a function of energy and time, with particular attention to cases
involving multiple crossing along the shock profile. Illustrative
applications are given in terms of positron spectra generated by
supernova electron antineutrinos through inverse beta decay.
\end{abstract}
\medskip
 \maketitle

\section{Introduction}

The realization that the supernova shock propagation can affect
neutrino flavor transitions \cite{Schi} a few seconds after core
bounce \cite{Wils} is gaining increasing attention in the recent
literature on supernova neutrinos
\cite{Raff,Beac,Lead,Reli,Spin,Shoc,Chou,Luna}. Indeed,
time-dependent variations of the neutrino potential in matter
\cite{Matt} along the supernova shock profile can leave an
interesting imprint on the energy and time structure of the
(anti)neutrino signal at the Earth \cite{Schi,Shoc,Luna}.
Observation of such possible effects, although very challenging
from an experimental viewpoint, would open a unique opportunity to
study some aspects of the supernova shock dynamics in ``real
time.'' Moreover, the strong sensitivity of such effects to the
neutrino oscillation parameters \cite{Schi,Shoc,Luna} might, in
principle, provide us with additional constraints (or hints, at
least) on the neutrino mass spectrum and mixing angles.

From a conservative viewpoint, it should be stressed that current
simulations of core-collapse supernovae, despite remarkable
efforts, may still require substantial physics improvements
\cite{Jank}. It is not excluded that more refined simulations
might significantly modify, e.g., the main features of shock
profile discussed in \cite{Schi}, and especially its propagation
velocity and density gradient, which govern the structure of the
neutrino crossing probabilities \cite{Schi,Shoc,Luna}. Moreover,
the time and energy dependence of the source (anti)neutrino fluxes
might be more complicated and uncertain \cite{Unce,Flux} than is
customarily assumed, making it more difficult to identify
shock-related signals. All these potential systematic
uncertainties in the basic physics ingredients might be large
enough to weaken the significance of shock-related signals, even
with hypothetically large experimental statistics. However,
despite all these caveats, the stakes are so high that further
investigations on possible shock effects on supernova neutrinos
are widely justified, in our opinion.

The purpose of this work is to investigate in some detail the
energy- and time-dependence of the neutrino crossing probabilities
along the supernova shock. We assume the active $3\nu$ oscillation
scenario (Sec.~II), and construct an empirical parameterization
for the shock profile (Sec.~III). Analytical calculations are
shown to reproduce well (and to be much more convenient than)
numerical approaches to the overall crossing probabilities, and
are used to break down the multiple transition structure along the
shock density profile, both in the energy domain (Sec.~IV) and in
the time domain (Sec.~V). Finally, applications are given in terms
of antineutrino event spectra observable at the Earth through
inverse beta decay (Sec.~VI). Conclusions and prospects for
further work are given in Sec.~VII.

The bottom line of our work is that simple, analytical
calculations of the crossing probabilities along the supernova
shock profile appear to be adequate to study the structure of
neutrino flavor transitions in the energy and time domain, for all
practical purposes. The specific probability values estimated in
this work, however, should not be taken too literally, being based
on a simplified description of the matter density profile, which
is subject to change as more detailed supernova shock simulations
are performed and become publicly available.

\section{$3\nu$  notation and analytical approximations}

Throughout this work, we consider only flavor transitions among
the three known active neutrinos $\nu_e$, $\nu_\mu$, and
$\nu_\tau$. In the following, we set the notation and briefly
describe the analytical expressions for the crossing
probabilities.

\subsection{Notation for $3\nu$ kinematics and dynamics}

The $3\nu$ squared mass spectrum is parameterized as
\begin{equation}
\label{mass2} (m^2_1,\,m^2_2,\,m^2_3)=\left(-\frac{\delta
m^2}{2},\,+\frac{\delta m^2}{2},\,\pm\Delta m^2\right)\ ,
\end{equation}
where the sign before $\Delta m^2$ distinguishes the cases of
normal ($+$) and inverted ($-$) hierarchy. The mixing matrix $U$
is defined in terms of three rotation angles $\theta_{ij}$,
ordered as for the quark mixing matrix \cite{Hagi},
\begin{equation}
\label{Umatr} U=U(\theta_{12},\,\theta_{13},\,\theta_{23}).
\end{equation}
When needed, we take
\begin{eqnarray}
\label{LMAm} \delta m^2/\mathrm{eV}^2 &=& 7.3\times 10^{-5}\ ,\\
\label{LMAs} \sin^2\theta_{12} &=& 0.315\ ,
\end{eqnarray}
corresponding to the so-called LMA-I best fit to the solar and
reactor neutrino data \cite{LMAI}. We also take
\begin{equation}
\label{K2Km} \Delta m^2=3\times 10^{-3}\mathrm{\ eV}^2\ ,
\end{equation}
close to the best fit to the atmospheric and accelerator neutrino
data detected in Super-Kamiokande \cite{Deco}. The value of
$\sin^2\theta_{23}$ is basically irrelevant for supernova
neutrinos, since it corresponds to an unobservable rotation in the
$(\nu_\mu,\nu_\tau)$ flavor subspace. Concerning the mixing
parameter $\sin^2\theta_{13}$, we will pick up several
representative values, well below the current $3\sigma$ upper
limit ($\sin^2\theta_{13}<0.05$ \cite{Deco}).

The mass gap hierarchy  ($\delta m^2/\Delta m^2\ll 1$) and the
smallness of $\sin^2\theta_{13}$ guarantee, to a very good
approximation, the factorization of the $3\nu$ neutrino dynamics
into a ``low'' ($L$) and a ``high'' ($H$) $2\nu$ subsystem, whose
dominant oscillation parameters are $(\delta m^2,\,\theta_{12})$
and $(\Delta m^2,\,\theta_{13})$, respectively (see
\cite{KuPa,Ours} and references therein). The corresponding
neutrino wavenumbers are defined as
\begin{eqnarray}
\label{kL} k_L &=& \delta m^2/2E\ ,\\
\label{kH} k_H &=& \Delta m^2/2E\ ,
\end{eqnarray}
where $E$ is the neutrino energy.

Potentially large matter effects \cite{Matt} are expected when
either $k_L$ or $k_H$ take values close to the neutrino potential
$V$ in matter,
\begin{equation}
\label{V} V=\sqrt{2}\, G_F\, N_e(x)\ ,
\end{equation}
where $N_e$ is the electron density at the supernova radius $x$.
In appropriate units,
\begin{equation}
\label{Vunits} \frac{V(x)}{\mathrm{eV}^2/\mathrm{MeV}} = 7.6\times
10^{-8} \,Y_e\, \frac{\rho(x)}{\mathrm{g/cm}^3}\ ,
\end{equation}
where $Y_e$ is the electron/nucleon number fraction (that we
assume equal to 1/2), while $\rho(x)$ is the radial mass density
profile. We conventionally take $k_L$, $k_H$, and $V$ to be
positive (as in Ref.~\cite{Ours}), irrespective of the neutrino
type ($\nu$ or $\overline\nu$) and hierarchy (normal or inverted).

The $V(x)$ profile in supernovae is often approximated by a static
power law, $V(x)\sim x^{-n}$ with $n\sim 3$. In the presence of a
shock wave, however, such static profile can be significantly
modified \cite{Schi,Wils}. In particular the shock wave, while
propagating outwards at supersonic speed, leaves behind a
rarefaction zone, followed by a high-density region and by a sharp
drop of density (down to the static value) at
\begin{equation}\label{xspos}
  x_s = \mathrm{shock\ front\ radius}\ .
\end{equation}
This abrupt change of density occurs over a microscopic length
scale governed by the mean free path of ions and electrons,
which, for our purposes, can be effectively taken as zero.%
\footnote{The authors of \cite{Schi} correctly recognized that the
apparent smoothness of numerical shock front profiles
\cite{Wils,Shoc} is actually an artifact of hydrodynamical
simulations, which tend to widen the true (step-like) profile over
several numerical grid zones.}
Correspondingly, at $x=x_s$ the matter potential $V(x)$ drops from
\begin{equation}\label{V+}
 V^+= \lim_{x\to x_s^-} V(x)
\end{equation}
to
\begin{equation}\label{V-}
 V^-= \lim_{x\to x_s^+} V(x)\ ,
\end{equation}
with a typical ratio \cite{Schi}
\begin{equation}\label{xiratio}
  \xi=\frac{V^+}{V^-}\simeq 10\ .
\end{equation}
Given the peculiar step-like features of the matter potential at
the shock front, in the following we discuss the analytical
expressions of the crossing probabilities for $x\neq x_s$ and
$x=x_s$ separately. We then study the general case of multiple
crossings along the neutrino trajectory.

\subsection{Crossing probability at $x\neq x_s$}

As previously mentioned, potentially large matter effects are
expected when $k_{L,H}\sim V$. In particular, if the condition
\begin{equation}
\label{V=k} V(x_H)=k_H
\end{equation}
occurs at some point $x_H (\neq x_s)$, there is a finite
probability $P_H$ of crossing between effective mass eigenstates
in the $H$ subsystem, which is given to a good approximation by
the so-called double exponential formula \cite{Petc}
\begin{equation}
\label{PH} P_H=\frac{\exp(2\pi r_H k_H
\cos^2\theta_{13})-1}{\exp(2\pi r_H k_H )-1}\
\end{equation}
for both neutrinos and antineutrinos (see \cite{Ours} and related
bibliography therein), with
\begin{equation}
\label{rH} r_H = \left|\frac{d\ln V(x)}{dx}\right|^{-1}_{x=x_H}\ .
\end{equation}
For all the $V(x)$ profiles and $\sin^2\theta_{13}$ values that
will be used in the following, the above formula reduces to the
well-known Landau-Zener (LZ) limit
\begin{equation}
\label{LZ} P_H\simeq \exp(-2\pi r_H k_H \sin^2\theta_{13})\ ,
\end{equation}
at any $\nu$ energy of practical interest.

In principle, the occurrence of the condition $k_L=V(x_L)$ at some
radius $x_L$ could also induce neutrino level crossing in the $L$
subsystem with probability $P_L$ (generally different for
neutrinos and antineutrinos; see, e.g., \cite{Digh,Ours}).
However, for $\sin^2\theta_{12}$ as large as in Eq.~(\ref{LMAs})
it turns out that the $L$-transition is adiabatic,
\begin{equation}
\label{PL} P_L(\nu)\simeq 0\simeq P_L(\overline\nu)\ ,
\end{equation}
at all points $x\neq x_s$.

\subsection{Crossing probability at $x= x_s$}

The analytical approximations in Eqs.~(\ref{LZ}) and (\ref{PL})
break down at  $x=x_s$, where the density gradient explodes (and
$r_H\to 0$). In this case, the crossing probability $P_s$ is
determined by the conservation of flavor across the $V(x)$
discontinuity, with the well-known result (see, e.g.,
\cite{KuPa}):
\begin{equation}\label{Ps}
  P_s=\sin^2(\theta_m^+-\theta_m^-)\ ,
\end{equation}
where $\theta_m^+$ and $\theta_m^-$ are the effective mixing
angles in matter immediately before $(V=V^+)$ and after $(V=V^-)$
the shock front position $x_s$, respectively. For the $H$
transition, such angles are defined by
\begin{equation}\label{theta+-}
  \cos 2\theta_m^\pm=\frac{\cos2\theta_{13}-V^\pm/k_H}
  {\sqrt{(\cos2\theta_{13}-V^\pm/k_H)^2+(\sin2\theta_{13})^2}}\ ,
\end{equation}
for both neutrinos and antineutrinos.%
\footnote{We remind the reader that, as proven in \cite{Ours},
$P_H(\nu)=P_H(\overline\nu)$, independently of the functional form
of $P_H$.}
In the limit of very small $\theta_{13}$, the above equations
imply $P_s\simeq 1$ for $k_H\in [V^-,V^+]$, and $P_s\simeq 0$
otherwise. Notice that the extremely nonadiabatic limit $P_s\simeq
1$  would also be obtained from Eq.~(\ref{PH}) for $\theta_{13}\to
0$ and $r_H\to 0$. However, the tempting approximation $P_s=1$
(for $k_H\in[V^-,V^+]$)  does not work well for
$\sin^2\theta_{13}$ as large as $O(10^{-3})$--$O(10^{-2})$, the
``top-hat'' behavior for $P_s$ being significantly smeared out
[i.e., a tail $P_s\neq 0$ is developed for $k_H$ outside the range
$[V^-,V^+]$, especially for $k_H\Lesssim V^-$, as easily
understandable on the basis of Eq.~(\ref{theta+-})]. Therefore, we
adopt the exact Eqs.~(\ref{Ps}) and (\ref{theta+-}) to calculate
the $H$ crossing probability at the shock front, without further
approximations.

Concerning the analogous crossing probability in the $L$ sector,
similar expressions apply, modulo the substitutions
$\theta_{13}\to\theta_{12}$ and $k_H\to k_L$ in Eqs.~(\ref{Ps}) and
(\ref{theta+-}).%
\footnote{Contrary to the $H$-transition case, it is $P_L(\nu)\neq
P_L(\overline\nu)$ in general. $P_L(\overline\nu)$ can be obtained
from $P_L(\nu)$ through the further substitution $V^\pm/k_L\to
-V^\pm/k_L$ \cite{Ours}.}
The strong nonadiabatic character of the transition at the shock
front can now lead to $P_L\neq 0$ in some corners of the parameter
space [in contrast to Eq.~(\ref{PL})]. However, as we shall
discuss at the end of Sec.~IV~C, such parameter values happen to
have little phenomenological relevance. Therefore, unless
otherwise noted, we shall formally take $P_L\simeq 0$ not only for
$x\neq x_s$ [Eq.~(\ref{PL})], but also for $x=x_s$.

\subsection{Analytical expressions for multiple crossings}

For $P_L\simeq 0$, the formal expression for the $\nu_e$ survival
probability $P_{ee}$
\cite{Digh,Ours} is exceedingly simple (up to Earth matter effects):%
\footnote{Examples of Earth matter effects will be illustrated in
Sec.~VI}
\begin{equation}
\label{cases} P_{ee} \simeq  \left\{
\begin{array}{ll}
 \sin^2\theta_{12}\, P_H & (\nu,\;\mathrm{normal}), \\
 \cos^2\theta_{12}       & (\overline\nu,\;\mathrm{normal}), \\
 \sin^2\theta_{12}       & (\nu,\;\mathrm{inverted}), \\
 \cos^2\theta_{12}\, P_H & (\overline\nu,\;\mathrm{inverted}), \\
\end{array}\right.
\end{equation}
where ``normal'' and ``inverted'' refer to the hierarchy type. In
the above equation, we  have neglected small additive terms of
$O(\sin^2\theta_{13})$, which are not relevant for our discussion.
From the above equations, it appears that $P_H$ can modulate the
(otherwise constant) survival probability of $\nu_e$ in normal
hierarchy and of $\overline\nu_e$ in inverted hierarchy, thus
providing an important handle to solve the current hierarchy
ambiguity.

In the presence of a propagating shock wave, the calculation of
$P_H$ is not entirely trivial. For the shock profiles studied in
\cite{Schi,Shoc}, the condition in Eq.~(\ref{V=k}) can occur at up
to three points $x_i\neq x_s\ (i=1,2,3)$:
\begin{equation}\label{Vk}
V(x_i)=k_H\ (i=1,2,3)\ ,
\end{equation}
In the most general case, two such points belong to the
rarefaction zone ($x_1 < x_2<x_s$) and one to the static region
below the shock front $(x_3>x_s)$. The corresponding crossing
probabilities are denoted as $P_1$, $P_2$, and $P_3$. The $P_i$'s
can be analitically calculated as in Eq.~(\ref{LZ}), with the
inverse log-derivative $r_i$ [Eq.~(\ref{rH})] evaluated at the
point $x_i$. In addition, one should consider the crossing
probability $P_s$ at $x=x_s$, as given by Eqs.~(\ref{Ps}) and
(\ref{theta+-}). Therefore, the global $P_H$ must be generally
constructed in terms of four crossing probabilities $P_1$, $P_2$,
$P_s$, and $P_3$, occurring at the points
\begin{equation}
\label{x123} x_1 < x_2 < x_s < x_3\ ,
\end{equation}
respectively.

If we assume that all relative neutrino phases can be averaged
away, and that the four transitions can be exactly factorized, the
overall crossing probability $P_H$ can be defined in terms of a
simple matrix equation \cite{KuPa}
\begin{equation}
\label{Matr} \left(
\begin{array}{cc}
1-P_H  & P_H \\
P_H & 1-P_H
\end{array}
\right)=  \prod_{i=1,2,s,3} \left(
\begin{array}{cc}
1-P_i  & P_i \\
P_i & 1-P_i
\end{array}
\right)\ ,
\end{equation}
whose solution is
\begin{eqnarray}
 P_H&=&P_1+P_2+P_s+P_3-2(P_1P_2+P_1P_s+P_1P_3+P_2P_s+P_2P_3+P_sP_3)\nonumber\\
 \label{PH123}
&&+4(P_1P_2P_s+P_1P_2P_3+P_1P_sP_3+P_2P_sP_3)-8 P_1P_2P_sP_3\ .
\end{eqnarray}
We will refer to this simple equation for our analytical
calculations of $P_H$.

Although the smallness of $\theta_{13}$ implies that each of the
four transitions $P_{1,2,s,3}$ is (in general) well localized and
can thus be separated from the others, the intrinsic accuracy of
the factorization in Eq.~(\ref{Matr})  is not obvious a priori in
the whole relevant parameter space. There are zones of the shock
profile where two transition points can become very close and
eventually merge, or where the local density shape can be strongly
different from the static, quasi-cubic power law which was used in
\cite{Ours} to test the analytical recipe in of Eq.~(\ref{PH}).
Moreover, in the presence of multiple (interfering) transition
amplitudes, it is worth checking that reasonable phase averaging
effectively reproduces the incoherent factorization implicit in
Eq.~(\ref{Matr}).

For such reasons, we think it is useful to compare the analytical
calculations of $P_H$ [based on Eqs.~(\ref{rH}) and (\ref{LZ}) for
$x=x_{1,2,3}$, and on Eqs.~(\ref{Ps}) and (\ref{theta+-}) for
$x=x_s$] with the results of a numerical (Runge-Kutta) evolution
of the neutrino flavor propagation equations along representative
shock density profiles. To our knowledge, such reassuring
comparison has not been explicitly performed in the available
literature.%
\footnote{The authors of Refs.~\cite{Shoc} and \cite{Schi} seem to
have used a fully numerical and a mixed numerical+analytical
approach to the neutrino evolution equations, respectively. The
authors of Ref.~\cite{Luna} discuss only qualitatively the
analytic form of $P_3$, in the hypothesis that $P_1\sim0\sim
P_2$.}
In order to do so in both the time and energy domain (Secs.~IV and
V, respectively), we need a parameterization of the shock wave
profile which is (almost everywhere) continuous in both the radius
$x$ and in post-bounce time $t$. Unfortunately, a similar
parameterization (or, equivalently, a dense numerical table) is
not publicly available; only some representative profiles are
graphically
reported in \cite{Schi} and \cite{Shoc}.%
\footnote{Notice that the bilogarithmic scales used in
\cite{Schi,Shoc} make it difficult to recover potentially
interesting shock details by graphical reduction.}
Moreover, only the profiles in \cite{Schi} exhibit (by
construction) the correct step-like profile at $x=x_s$. In the
following section, we thus introduce a simplified parametrization
of the supernova density profile which is continuous in the
$(x,t)$ variables [excepting the shock-front discontinuity point].
Such empirical profile captures the main qualitative features of
the time and radial dependence of the (supposed spherically
symmetric) shock propagation, allowing us to perform a meaningful
comparison of numerical and analytical results, and to break down
explicitly the analytical calculation of $P_H$ into its four
components $P_1$, $P_2$, $P_s$, and $P_3$. Needless to say, such a
simplified profile can and must be improved when more informative
or refined supernova shock simulations become available.

A final remark is in order. In this work we are mainly concerned
with the factorization of transition {\em within\/} the $H$
subsystem, since the factorization {\em between\/} the $L$ and $H$
subsystems is guaranteed for $\delta m^2$ as low as in
Eq.~(\ref{LMAm}). However, small corrections to the $L$-$H$
factorization might arise for values of $\delta m^2 \sim
\mathrm{few}\times 10^{-4}$ eV$^2$, not considered here (see,
e.g., \cite{KuPa} for earlier discussions and \cite{Luna,LimS} for
more recent considerations). The analysis of these  possible small
corrections is beyond the scope of this work.

\section{Empirical parametrization of the Shock density profile}

In this section we introduce a simplified, empirical
parametrization of the shock density profile $\rho$, which
reproduces the main features of the graphical profile in
\cite{Schi} and, at same time, is continuous in both $x$ and $t$
[excepting the case $x(t)=x_s(t)$]. In this way, numerical and
analytical calculations of $P_H$ can continuously cover the
relevant parameter space.

For post-bounce times $t\Lesssim 1\,\mathrm{s}$, shock effects
\cite{Schi,Shoc} take place typically at $V > k_H$ and do not
significantly affect the $H$ subsystem. The density profile can
thus be effectively approximated by its static limit $\rho_0$ as
taken from \cite{Schi}:
\begin{equation}
\label{static} t\Lesssim 1\,\mathrm{s}\ \ \Rightarrow\ \
\frac{\rho_0(x)}{\mathrm{g/cm}^3}\simeq
10^{14}\,\left(\frac{x}{\mathrm{km}}\right)^{-2.4}\ .
\end{equation}

For post-bounce times $t\Gtrsim 1\,\mathrm{s}$, the condition $V
\sim  k_H$ can be fulfilled, and the shock profile can thus
modulate $P_H$. We characterize the profile in terms of the shock
front position $x_s$ and of its shape variation $f(x)$ (with
respect to $\rho_0$) for $x\leq x_s$. Formally, one can write
\begin{equation}
\label{profile} t\Gtrsim 1\,\mathrm{s}\ \ \Rightarrow\ \
\rho(x)=\rho_0(x)\cdot\left\{
\begin{array}{ll}
\xi\cdot f(x) &,\;x\leq x_s\ , \\
1        &,\;x>x_s    \ ,
\end{array}\right.
\end{equation}
where $\xi$ (the matter density enhancement at $x=x_s$) has been
defined in Eq.~(\ref{xiratio}) (see also \cite{Schi}).

In Eq.~(\ref{profile}), the function $f(x)$ parametrizes the
rarefaction zone (``hot bubble'') above the shock front,
characterized by a drop of density over more than one decade in
$x$, and by an asymptotic density increase for $x\ll x_s$. After
some trials, we have chosen the following (purely empirical)
parametrization for $f(x)$, which reproduces the main features of
the graphical profiles in \cite{Schi}:
\begin{equation}
\label{f<} \ln f(x)
=\left[0.28-0.69\ln(x_s/\mathrm{km})\right]\cdot
\left[\arcsin(1-x/x_s)\right]^{1.1}\ ,
\end{equation}

In the above equations we assume (following \cite{Schi}) a
slightly accelerating shock-front position $x_s$,
\begin{equation}
\label{accel} x_s(t)=x^0_s+v_s\,t+\frac{1}{2}\, a_s\, t^2\ ,
\end{equation}
with parameters approximately given by
\begin{eqnarray}
\label{x0} x^0_s&\simeq& -4.6 \times 10^3\mathrm{\ km} \ , \\  %
\label{vs} v_s  &\simeq& 11.3 \times 10^3\mathrm{\ km/s}\ , \\ %
\label{as} a_s  &\simeq&  0.2 \times 10^3\mathrm{\ km/s}^2\ .
\end{eqnarray}
Notice that: ($i$) The time dependence of $\rho(x)$ in
Eq.~(\ref{profile}) is implicit, through the function
$x_s=x_s(t)$; and ($ii$) at small radii, it is
\begin{equation}
\label{x<} x\ll x_s \ \ \Rightarrow\ \ \rho(x) \simeq \rho_0(x)
\times 16\,(x_s/\mathrm{km})^{-1.13} \ .
\end{equation}

\begin{figure}
\vspace*{-3.6cm}\hspace*{-2.6cm}
\includegraphics[scale=0.92, bb= 30 100 500 700]{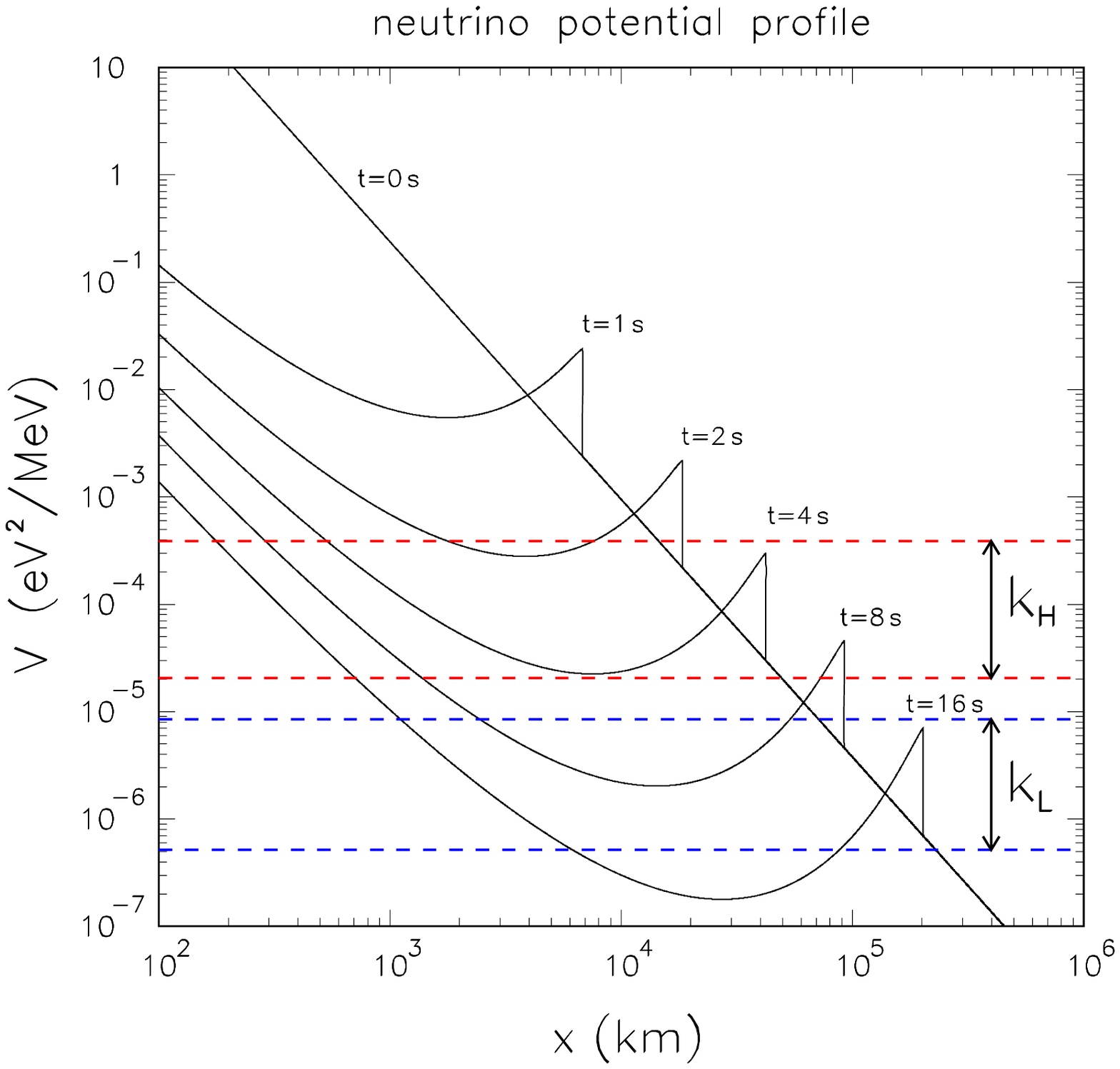}
\vspace*{2.6cm} \caption{\label{fig01} Neutrino potential $V$ as a
function of the supernova radius $x$, for different values of the
post-bounce time $t$, as derived from our simplified
parametrization of the profiles in \cite{Schi}.  The horizontal
bands represent typical ranges of the neutrino wavenumbers $k_H$
and $k_L$. See the text for details.}
\end{figure}

Figure~\ref{fig01} shows the neutrino potential $V(x)$ as derived
from the above parametrization [Eqs.(\ref{static})--(\ref{as})]
for representative post-bounce times  $\geq 1$~s, as well as for
the static profile $(t\simeq 0)$. These curves reasonably
reproduce the main features of the shock profiles from
\cite{Schi}. The horizontal bands represent the ranges where the
neutrino potential equals the two wavenumbers ($k_H=V$ and
$k_L=V$), for a representative energy interval $E\in[4,\,70]$~MeV.

As previously mentioned, a line at constant $k_H$ can intersect
the $V(x)$ profile in Fig.~\ref{fig01} in at most three points
$x_i$, ordered as $ x_1 \leq x_2\, (\leq x_s) \leq x_3$, and
corresponding to level crossing probabilities $P_1$, $P_2$, and
$P_3$. A further crossing probability $P_s$ is related to the
step-like feature of the profile at the shock front $x=x_s$. Two
of the four critical points ($x_1,x_2,x_s,x_3$) can merge in the
following cases: (1) at the bottom of the rarefaction zone
(denoted as $x_r$), where $x_1=x_2$; (2) at the shock front
$(x_s)$, where it is $x_2=x_s$ for $V^+=k_H$ or $x_3=x_s$ for
$V^-=k_H$. Single transitions can instead occur at early times or
high energies ($V(x)=k_H$ at $x=x_3$ only), as well as at late
times or low energies ($V(x)=k_H$ at $x=x_1$ only). Therefore, the
three critical cases $V(x_r)=k_H$ ($x_1=x_2$) and $V^\pm=k_H$
($x_{2,3}=x_s$) are expected to mark significant changes in the
behavior of $P_H$.

We conclude this section with a few cautionary remarks. At
present, the detailed shape of $V(x)$ at the shock front is poorly
known, since the physical requirement of a density discontinuity
at $x=x_s$ is implemented ``by hand'' \cite{Schi}, as a necessary
correction to the artificially smooth profile from simulations
\cite{Shoc}.%
\footnote{In Ref.~\cite{Schi}, the authors  explicitly say to have
steepened the shock front because ``... in hydrodynamics
calculations [it] may be softened by numerical techniques.''}
A side effect of the shock front steepening is the appearance of a
local ``cusp'' at $x=x_s$ (as apparent in Ref.~\cite{Schi} and
even more in our Fig.~\ref{fig01}), which might well be
unphysical. However, we do not smooth it out, since it actually
provides us with a useful, nontrivial check of numerical versus
analytical calculations of $P_H$ in an ``extreme'' condition
(i.e., around a sudden change in the sign and value of the density
gradient). Finally, we observe that some ``leftover'' turbulence
might be expected in the rarefaction zone behind the shock front,
inducing fuzzy variations of the local density. Such variations
might lead to more than three solutions of Eq.~(\ref{Vk}) in some
cases, and thus to a more complicated and ``random'' structure for
$P_H$. This possibility (not considered in this work) should be
kept in mind when more refined supernova shock simulations will
become available.

\section{Analysis of crossing probabilities in the energy domain}

In this Section we study in detail the energy dependence of $P_H$
at fixed time. We pay particular attention to the discussion of
multiple transitions along the shock density profile.

\subsection{Analytical {\em vs\/} numerical calculations of $P_H(E)$}

\begin{figure}
\vspace*{+2.5cm}\hspace*{-2.4cm}
\includegraphics[scale=0.85, bb= 30 100 500 700]{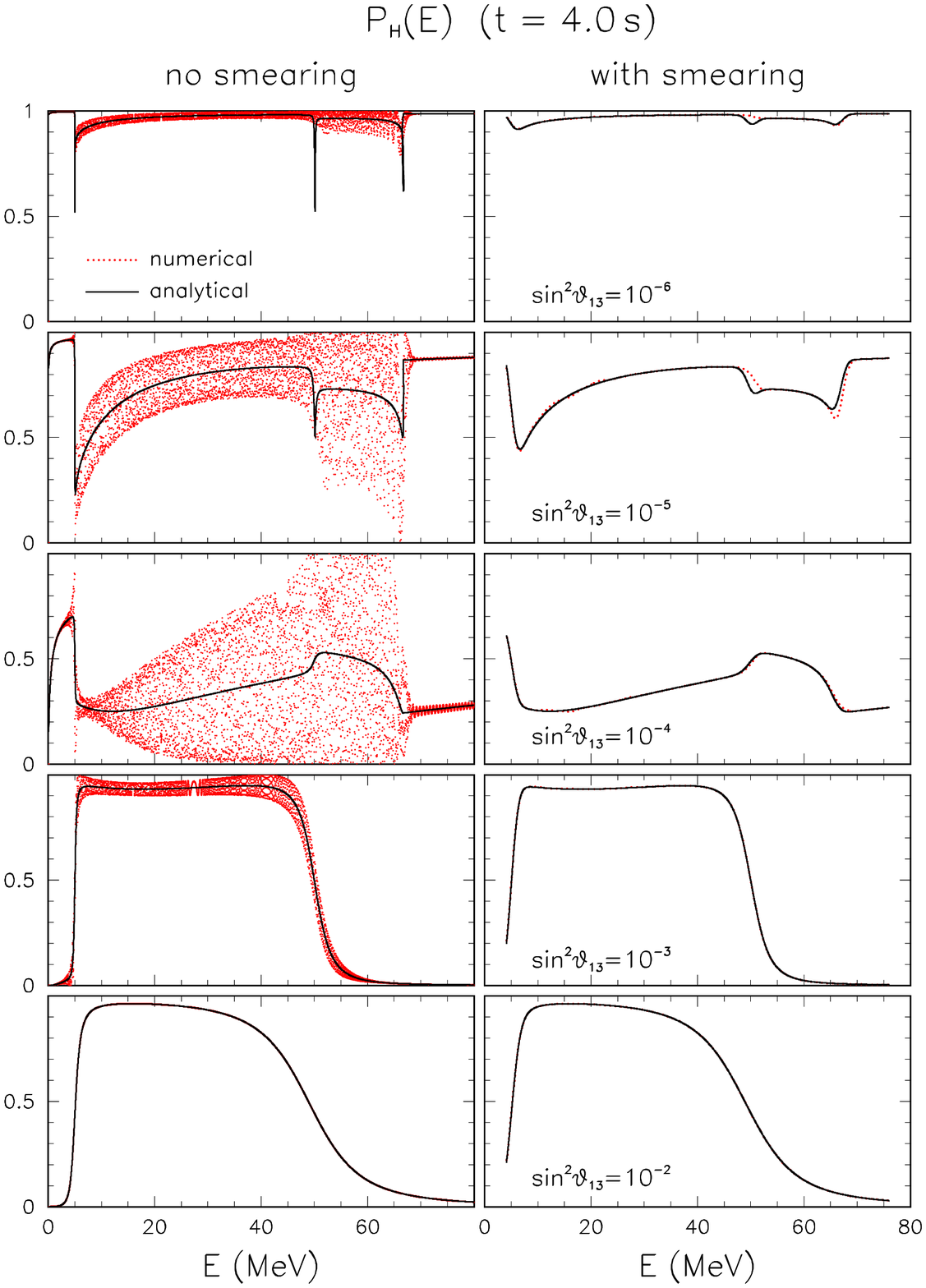}
\vspace*{.6cm} \caption{\label{fig02} Comparison of numerical and
analytical calculations of the crossing probability $P_H(E)$ (dots
and solid curves, respectively) at $t=4\,\mathrm{s}$ and for five
representative values of $\sin^2\theta_{13}$. In the right panels,
phase effects are energy-averaged.}
\end{figure}

Figure~\ref{fig02} shows our calculation of $P_H(E)$ at fixed
post-bounce time $t=4\,\mathrm{s}$, for five representative values
of $\sin^2\theta_{13}$, ranging from $10^{-6}$ (top) to $10^{-2}$
(bottom). In the left panels, a direct comparison is made between
analytical calculations (solid curves) and Runge-Kutta numerical
calculations (dots), performed at $5\times 10^3$ equally-spaced
points in the interval $E\in[0,80]$ MeV. In general, up to four
energy regimes can be identified, with significant changes at the
three critical energies $E\simeq 5$, 50, and 67 MeV. These
energies fulfill (for the $V(x)$ profile at $t=4$~s) the three
critical conditions $k_H=V^+,V^-,V(x_r)$ which, as discussed in
the previous section, signal the merging of two of the four
possible level crossing points.%
\footnote{Note that these critical conditions are independent of
$\theta_{13}$.}

In the left panel of Fig.~\ref{fig02}, the results of the
numerical calculations appear to be generally scattered around the
analytical curve. The reason is that Runge-Kutta calculations keep
track of the phase(s) of the level crossing amplitude(s), while
our analytical approximations average out this information from
the beginning. For a single numerical amplitude, the associated
phase becomes ineffective after taking the squared modulus. For
multiple (interfering) amplitudes, instead, the relative phases do
provide an oscillating structure in $P_H$, which is apparent in
the numerical results.%
\footnote{In the left panels of Fig.~\ref{fig02}, for the sake of
graphical clearness, we show the spread of numerical results as
scatter plots, rather than through rapidly oscillating curves.}

In the right panels of Fig.~\ref{fig02}, the oscillating structure
of the numerical results is averaged out by a convolution with a
Gaussian function (with one-sigma width of $\pm 1$ MeV), which
simulates a generic ``smearing'' process (e.g., experimental
energy resolution). The same convolution is applied to the
analytical results. It can be seen that there is very good
agreement between the two different (numerical and analytical)
approaches after smearing.  The results of Fig.~\ref{fig02} (and
of other checks that we have performed for different post-bounce
times) show that analytical calculations of $P_H$ coincide with
phase-averaged numerical results with very good accuracy, even
close to the ``critical'' points where two transitions collapse.
Moreover, analytical calculations are actually much more
convenient than the numerical ones, which require a very high
sampling rate in order to perform efficient phase-averaging and to
prevent numerical artifacts.

Let us now discuss in more detail the behavior of the analytical
$P_H$ in the left panels of Fig.~\ref{fig02}. We remind that $P_s$
[the crossing probability at the shock front, Eq.~(\ref{Ps})] has
almost a top-hat behavior: it rapidly drops from $\sim 1$ to zero
for $k_H$ outside the range $[V^-,V^+]$, i.e., for $E$ outside the
range $[5,50]$ MeV (for the specific profile used in
Fig.~\ref{fig02}). This behavior depends only mildly on
$\sin^2\theta_{13}$ [as far as it is $\Lesssim O(10^{-2})$]. The
LZ probabilities $P_{1,2,3}$ are instead exponentially suppressed
as $\sin^2\theta_{13}$ increases [see Eq.~\ref{PH}]. In
particular, in the bottom panels of Fig.~\ref{fig02}
($\sin^2\theta_{13}= 10^{-3}$ and $10^{-2}$), $P_H$ is dominated
by $P_s$ (with its characteristic top-hat behavior), with very
little contributions from $P_{1,2,3}$. The dominance of a single
crossing amplitude in $P_H$ explains the suppression of the
numerical oscillations in the same panels.

Conversely, for $\sin^2\theta_{13}$ as small as $10^{-6}$ eV$^2$
(top panel in Fig.~\ref{fig02}), the probabilities $P_{1,2,3}$ (as
well as $P_s$) can all become as large as $\sim 1$, rendering
$P_H\simeq 1$. In particular, for $E\Lesssim 5$ MeV, only
$P_1\simeq 1$ is active (single crossing, well above the shock
front). For $5\Lesssim E \Lesssim 50$ MeV, besides $P_s\simeq 1$
it is also $P_{1,2}\simeq 1$ (two crossings in the rarefaction
zone). For $50 \Lesssim E\Lesssim 67$ MeV, $P_{1,2}$ are still
active, $P_3\simeq 1$ is switched on, and $P_s$ drops to zero.
Finally, for $E\gtrsim 67$ MeV, only $P_3$ (level crossing well
below the shock front) survives. In all such cases (characterized
by either single or triple crossing with strong nonadiabatic
character) one derives from Eq.~(\ref{PH123}) that $P_H\simeq 1$.

The intermediate panels in Fig.~\ref{fig02}, corresponding to
$\sin^2\theta_{13}=10^{-5}$ and $10^{-4}$, are not as easily
understood as the previous ones, since at least one of the three
probabilities $P_{1,2,3}$ is definitely $<1$. In such cases, $P_H$
is better understood by explicitly separating its components
$P_{1,2,s,3}$, as done in the next subsection.

\subsection{$P_H(E)$ decomposition}

\begin{figure}
\vspace*{+2.4cm}\hspace*{-2.8cm}
\includegraphics[scale=0.75, bb= 30 100 500 700]{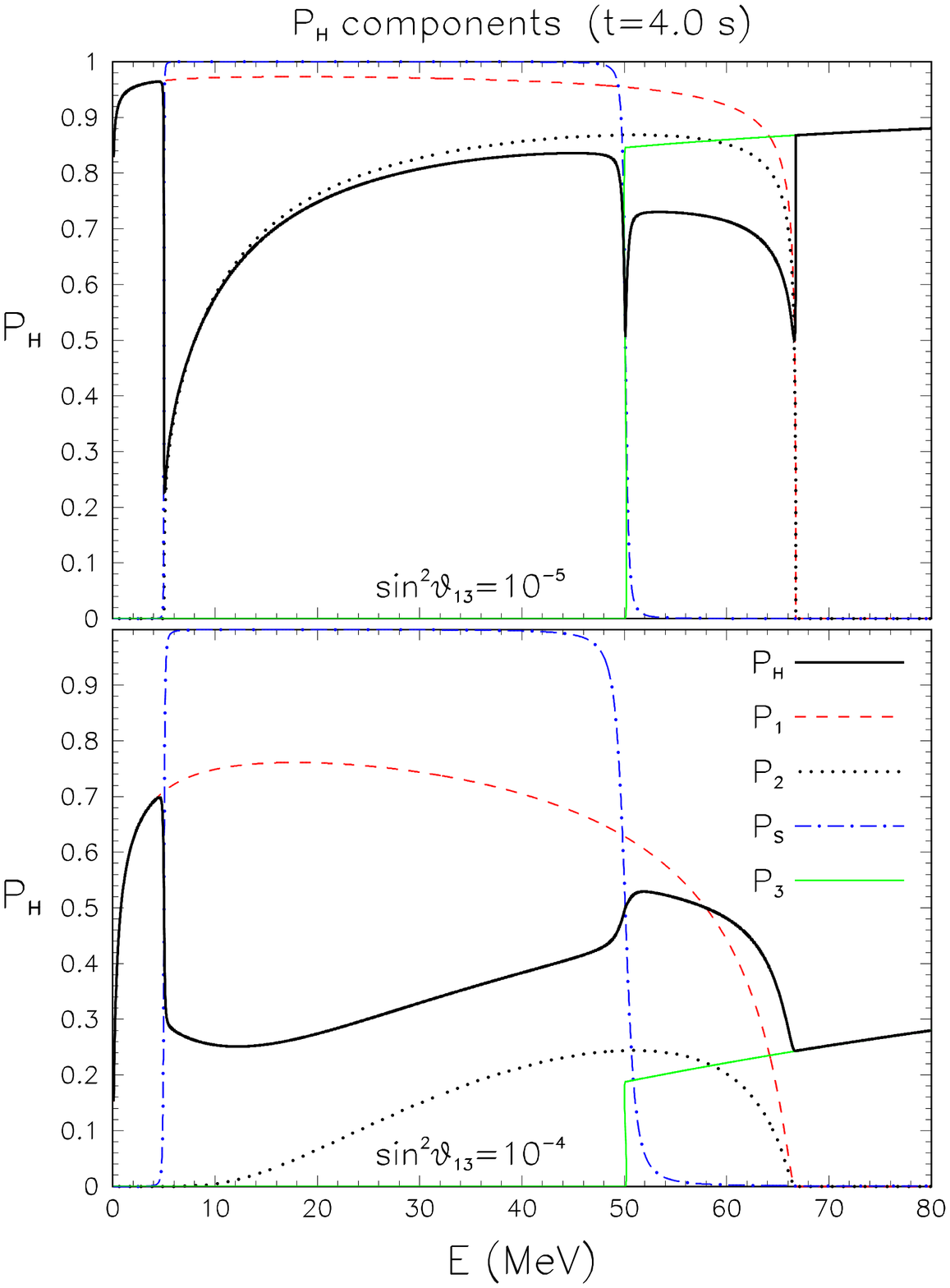}
\vspace*{2.cm} \caption{\label{fig03} $P_H(E)$ in terms of its
components $P_{1,2,s,3}$ at $t=4\,\mathrm{s}$, for two
representative values of $\sin^2\theta_{13}$.}
\end{figure}

Figure~\ref{fig03} shows the decomposition of $P_H(E)$
(analytical, unsmeared)  into the four crossing probabilities
$P_{1,2,s,3}$  for $\sin^2\theta_{13}=10^{-5}$ (top panel) and
$\sin^2\theta_{13}=10^{-4}$ (bottom panel). In both cases, the
shock wave profile at $t=4$~s (see Fig.~\ref{fig01}) is assumed.
The various transition probabilities are switched on and off at
the three critical energies ($E\simeq 5$, 50, and 67 MeV)
associated to such profile, as previously discussed.

In the top panel of Fig.~\ref{fig03}, both $P_1$ and $P_3$ are
strongly nonadiabatic ($\sim\!1$), and thus $P_H=P_1\sim 1$ and
$P_H=P_3\sim 1$ in the single-transition regimes at low and high
energy, respectively. At intermediate energies  both $P_s$ and
$P_2$ are nonzero, but the nonadiabatic character of $P_2$ is less
pronounced, since at $x_2$ (outer part of the rarefaction zone),
the density gradient is smaller [and the scale factor in
Eq.~(\ref{rH}) is larger] than at $x_1$ or $x_3$---at least in our
simplified description of the density profile. Therefore, in the
intermediate regime $5\Lesssim E \Lesssim 50$ MeV, the strong
nonadiabaticity of $P_1$ and $P_s$ implies $P_H\sim P_2$ through
Eq.~(\ref{PH123}). In summary, in the top panel of
Fig.~\ref{fig03}, the overall crossing probability $P_H$ initially
takes the value $P_1$, then $P_2$, and eventually $P_3$, as the
energy increases and passes through its  critical values. All
relevant parts of the density profile (static profile, shock
front, and both sides of the rarefaction zone) can thus contribute
to the $P_H$ energy profile.

In the bottom panel of Fig.~\ref{fig03}, the higher value of
$\sin^2\theta_{13}$ leads to an overall decrease of $P_{1,2,3}$,
particularly for the second transition ($P_2\sim 0$), while $P_s$
remains strongly nonadiabatic.  Therefore, the behavior of $P_H$
is not obvious, except at very low energy (where only $P_1>0$) or
at very high energy (where only $P_3>0$). We can only say that the
smallness of $P_2$ implies that the  outer part of the rarefaction
zone (just behind the shock front) are not relevant in this case.

Summarizing, variations of $\sin^2\theta_{13}$ for fixed density
profile can modulate significantly the $P_{1,2,3}$ contributions
to the crossing probability $P_H$. The contribution of $P_s$
remains instead stable and strongly nonadiabatic. As a
consequence, different parts of the neutrino profile can play
leading or subleading roles, according to the chosen value of
$\theta_{13}$. When $\theta_{13}$ will be known, one shall be able
to gauge the relative importance of such contributions, and to
determine which parts of the shock profile should be most
intensively studied. In the meantime, all parts of the shock
profile should be considered as potentially important and worth
further study.

\subsection{$P_H(E)$ at different times, and comments on $P_L$}

We conclude the analysis in the energy domain by briefly
discussing the variations of $P_H(E)$ for different post-bounce
times $t$. We also comment on the contribution of the shock-front
discontinuity to $P_L$.

\begin{figure}
\vspace*{+2.4cm}\hspace*{-2.8cm}
\includegraphics[scale=0.75, bb= 30 100 500 700]{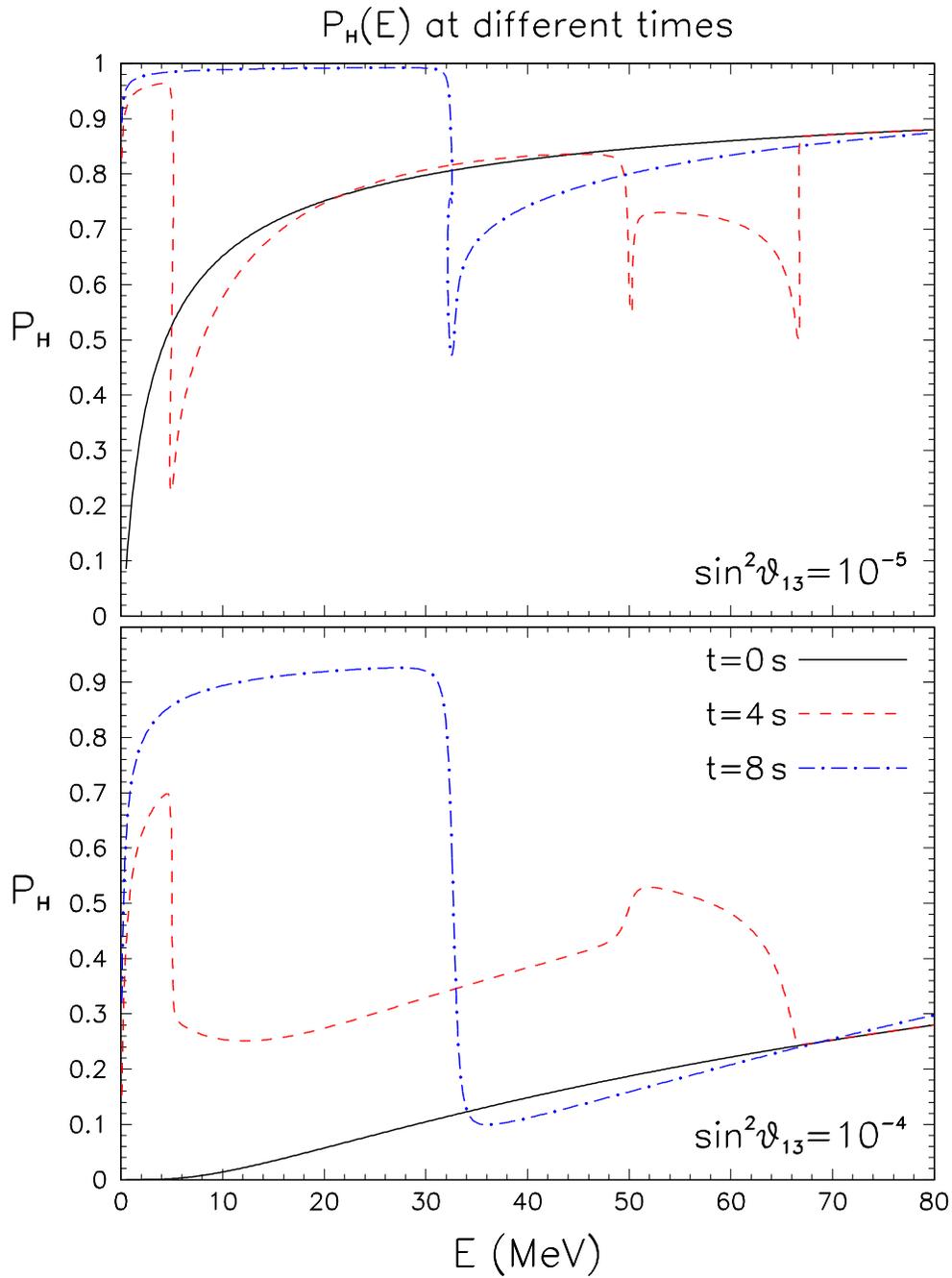}
\vspace*{2.cm} \caption{\label{fig04}  $P_H(E)$ at $t=0$, 4, and
$8\,\mathrm{s}$, for two representative values of
$\sin^2\theta_{13}$.}
\end{figure}

Figure~\ref{fig04} shows the function $P_H(E)$ at three different
times ($t=0$, 4, and 8~s) for $\sin^2\theta_{13}=10^{-5}$ (top
panel) and $10^{-4}$ (bottom panel). The curves at
$t=4\,\mathrm{s}$ have already been discussed. At earlier times
$t\Lesssim 1\,\mathrm{s}$ ($t=0\,\mathrm{s}$ in Fig.~\ref{fig04})
the behavior of $P_H$ becomes simpler, being basically dictated by
a single LZ crossing along the static profile. At later times
$(t=8\,\mathrm{s})$, the critical energies at which $k_H=V^\pm$
and $k_H=V(x_r)$ move forward and partly fall outside the energy
range relevant for supernova neutrinos.  As a consequence, the
nonadiabatic character of the innermost transition at $x_1$ makes
$P_H\sim 1$ in almost the first half of the energy range of
Fig.~\ref{fig04}, with a rapid drop to smaller values as the more
adiabatic transition $P_2$ becomes relevant. The drift of the
zones where $P_H$ is high or low has thus an interesting time
structure, which is analyzed in more detail in the next section.

We conclude this section with a few comments on the crossing
probability $P_L$ in the $L$ subsystem. As mentioned in Sec.~II~C,
the step-like nature of the shock front profile can make
$P_L\simeq P_s\neq 0$ for $k_L$ roughly within the range
$[V^-,V^+]$. In practice, however, this condition occurs only in
the low-energy tail of the (anti)neutrino spectrum $(E\Lesssim
\mathrm{few\ MeV})$ and at relatively late times $(t\Gtrsim
10\,\mathrm{s})$. Low-energy effects are suppressed by the lower
cross section and by experimental detection thresholds, while
late-time effects are intrinsically suppressed by the exponential
decrease of the supernova neutrino luminosity on a timescale of a
few seconds (see also Sec.~VI). Moreover, the $L$-transition is
never strongly nonadiabatic; in fact, from Eq.~(\ref{Ps}) one can
prove that $P_L(\nu)$ and $P_L(\overline\nu)$ are always
significantly smaller than $\cos^2\theta_{12}$ and
$\sin^2\theta_{12}$, respectively. For such reasons, effects
related to $P_L\neq 0$ at the shock front are hardly observable,
and have been neglected throughout this work.

\section{Analysis of crossing probabilities in the time domain}

In this Section we study in detail the time dependence of $P_H$ at
fixed energy. As in Sec.~IV, we start with a comparison between
numerical and analytical calculations of $P_H$, and then we study
the decomposition of $P_H$ in terms of $P_{1,2,s,3}$.

\subsection{Analytical {\em vs\/} numerical calculations of $P_H(t)$}

\begin{figure}
\vspace*{+2.5cm}\hspace*{-2.4cm}
\includegraphics[scale=0.85, bb= 30 100 500 700]{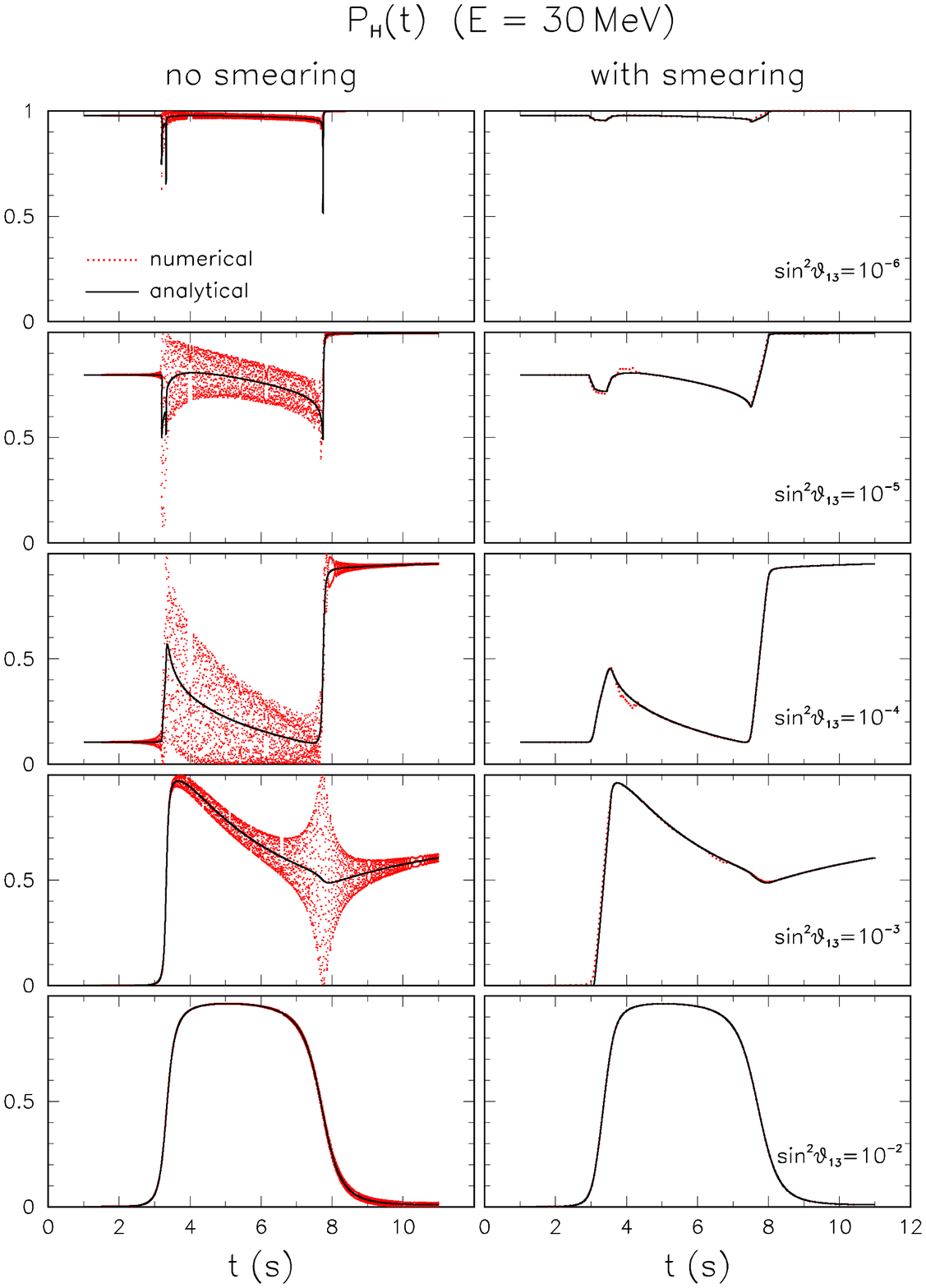}
\vspace*{.6cm} \caption{\label{fig05} Comparison of numerical and
analytical calculations of the crossing probability $P_H(t)$ (dots
and solid curves, respectively) at $E=30$~MeV and for five
representative values of $\sin^2\theta_{13}$. In the right panels,
phase effects are time-averaged.}
\end{figure}

Figure~\ref{fig05} shows our calculation of $P_H(t)$ at fixed
neutrino energy $E=30$ MeV and for five representative values of
$\sin^2\theta_{13}$, ranging from $10^{-6}$ (top) to $10^{-2}$
(bottom). In the left panels, a direct comparison is made between
analytical calculations (solid curves) and Runge-Kutta numerical
calculations (dots), performed at $5\times 10^3$ equally-spaced
points in the shown time interval. Once again, the interference
among multiple $\nu$ transition amplitudes appears to generate
fast oscillations, which scatter the numerical results around the
analytical curves. In the right panels of Fig.~\ref{fig05}, phase
averaging (smearing) is instead enforced through a convolution
with a ``top-hat'' time resolution function with $\pm
0.5\,\mathrm{s}$ width. It can be seen that such phase averaging
brings the numerical and analytical calculations in very good
agreement at any $t$, up to small residual artifacts in the
smeared numerical results, which would disappear by sampling the
abscissa more densely (not shown). From the results of
Figs.~\ref{fig02} and \ref{fig05} we conclude that simple,
analytical calculations provide the correct phase-averaged value
of $P_H$ as a function of both $E$ and $t$.  Therefore,
time-consuming numerical estimates of neutrino transitions along
the shock profile \cite{Schi,Shoc} can be accurately replaced by
much faster and elementary calculations. This is one of the main
results of our work.

 Let us now discuss in more detail the
behavior of the analytical $P_H$ in the left panels of
Fig.~\ref{fig05}. In analogy with the previous discussion in the
energy domain, the structure of $P_H$ appears to be characterized
by (up to) four different regimes also in the time domain. These
regimes are separated by three ``critical times,'' namely,
$t\simeq 3.2\,\mathrm{s}$ (when $k_H$ equals $V$ at the bottom of
the rarefaction zone $x_r$), $t\simeq 3.3\,\mathrm{s}$ (when $k_H$
equals $V^-$ at the bottom of the shock front), and $t\simeq
7.7\,\mathrm{s}$ (when $k_H$ equals $V^+$ at the top of the shock
front). [Of course, for $E\neq 30$~MeV such critical times will be
different.] The probabilities $P_{1,2,s,3}$ play different roles
in forming the global $P_H(t)$ in these different regimes. For
$t\Lesssim 3.2$~s, the relevant density profile is essentially
static, and $P_H\simeq
P_3\propto\mathrm{const}^{\sin^2\theta_{13}}$. Similarly, for
$t\Gtrsim 7.7\,\mathrm{s}$, it is $P_H\simeq
P_1\propto\mathrm{const}^{\sin^2\theta_{13}}$. The
single-transition values of $P_H$ at early and late times are thus
reduced by successive {\em powers\/} of ten as $\sin^2\theta_{13}$
is reduced by {\em factors\/} of ten from top to bottom in
Fig.~\ref{fig05}. At intermediate times the situation is instead
less obvious. For $t\simeq 3.2$--3.3~s, the crossing probabilities
$P_2$ and $P_s$ are switched on in succession, leading to a
complex local structure in $P_H(t)$ (especially at small
$\sin^2\theta_{13}$); in this situation, neutrino matter effects
are probing the lower part of both the rarefaction zone and of the
shock front (see Fig.~\ref{fig01}). In the interval $3.3\Lesssim
t\Lesssim 7.7$~s, $P_s$ is always strongly nonadiabatic, $P_3$ is
negligible, while $P_{1,2}$ play a role only at small
$\sin^2\theta_{13}$ (being exponentially suppressed for increasing
$\theta_{13}$). For $\sin^2\theta_{13}=10^{-2}$ (bottom panels),
only $P_s$ is important in forming the global $P_H$.

The most complicated cases in Fig.~\ref{fig05} thus emerge at
``intermediate times'' (i.e., when both the rarefaction zone and
the shock front are being probed) and at ``intermediate
$\theta_{13}$ mixing'' (i.e., when $P_{1,2,3}$ are moderately
nonadiabatic). In such cases, a decomposition of $P_H$ into
$P_{1,2,s,3}$ is necessary to grasp the main features of $P_H(t)$.

\subsection{$P_H(t)$ decomposition}

\begin{figure}
\vspace*{+2.4cm}\hspace*{-2.8cm}
\includegraphics[scale=0.75, bb= 30 100 500 700]{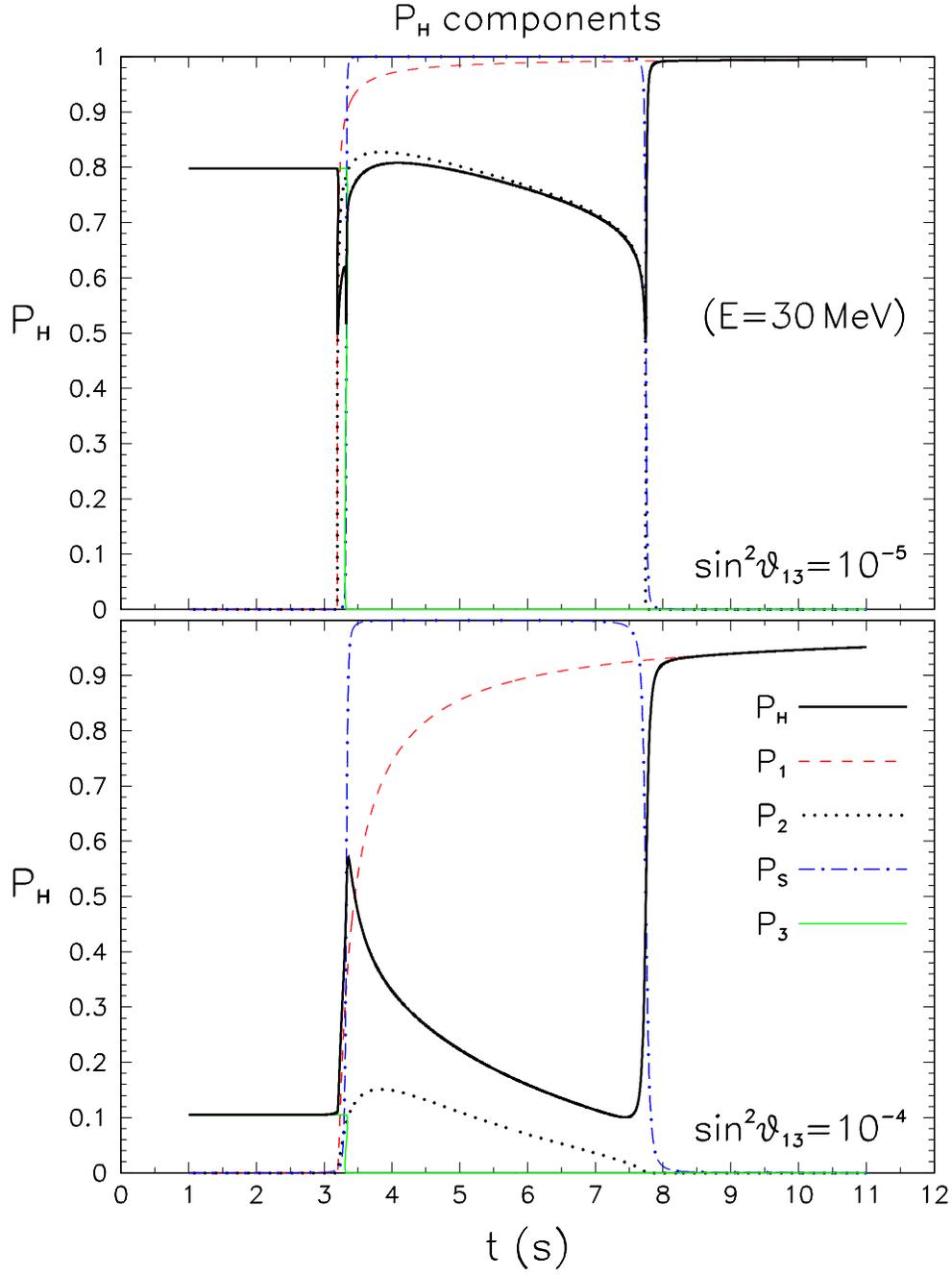}
\vspace*{2.cm} \caption{\label{fig06} $P_H(t)$ in terms of its
components $P_{1,2,s,3}$ at $E=30$ MeV, for two representative
values of $\sin^2\theta_{13}$.}
\end{figure}

Figure~\ref{fig06} shows the decomposition of $P_H(t)$
(analytical, unsmeared)  into the four crossing probabilities
$P_{1,2,s,3}$ at $E=30$ MeV, for $\sin^2\theta_{13}=10^{-5}$ (top
panel) and $\sin^2\theta_{13}=10^{-4}$ (bottom panel). The various
transition probabilities are switched on and off at the three
critical times $t\simeq 3.2$, 3.3, and 7.7~s. In both panels,
$P_3$ and $P_1$ dominate at early and late times, respectively. At
intermediate times, the structure of $P_H$ depends sensitively
[through Eq.~(\ref{PH123})] upon the specific values of the
crossing probabilities $(P_{1,2})$ in the rarefaction zone ($P_s$
being almost constant and close to $1$). The results in this
figure confirm that variations of $\theta_{13}$ can lead to
dramatic variations in the role of each $P_i$ (and thus of the
different parts of the shock profile) in building the final
crossing probability $P_H$ as a function of both energy and time.

\section{Impact on observable positron spectra}

As an application of the analytical calculation of $P_H$ described
in the previous section, we study the effect of the shock
propagation on the energy (and time) spectra of positrons
detectable at the Earth through the inverse beta-decay reaction
\begin{equation}
\overline \nu_e+p\to n+e^+\ .
\end{equation}
We assume a ``standard'' supernova explosion at $D=10$ kpc,
releasing a total energy $E_B=3\times 10^{53}$ erg, equally shared
among the (anti)neutrino flavors $\nu_\alpha$, and distributed in
time as
\begin{equation}\label{Lum}
L(t)=\frac{E_B}{6\tau}\,e^{-t/\tau}\ ,
\end{equation}
where $L$ is the luminosity for each $\nu_\alpha$ flux, and the
decay time is taken as $\tau=3\,\mathrm{s}$ \cite{Schi}. For the
sake of simplicity, we assume unpinched (normalized) Fermi-Dirac
$\nu_\alpha$ spectra with time-independent temperatures
$T_\alpha$,
\begin{equation}
\label{flux} f_\alpha(E) =
\frac{0.176}{T^4_\alpha}\,\frac{L(t)}{4\pi D^2}
\frac{E^2}{e^{E/T_\alpha}+1}\ ,
\end{equation}
where we take $T_{\overline e}=4.5$ MeV and $T_{\overline x}=6.5$
MeV ($x=\mu,\tau$).
\footnote{For  recent discussions of supernova neutrino energy
spectra, see \cite{Unce,Flux}.}

In the presence of oscillations, the supernova $\overline\nu_e$
spectrum is given by
\begin{equation}
\label{perm}
 f(E) = f_{\overline e}(E) P_{ee}  + f_{\overline
x}(E)(1-P_{ee}),
\end{equation}
where $P_{ee}$ is taken from Eq.~(\ref{cases}) (antineutrino
case). The convolution of $f(E)$ with the differential cross
section $d\sigma(E,E_\mathrm{pos})/dE_\mathrm{pos}$ \cite{Stru}
provides the event spectrum $dN/dE_\mathrm{pos}$ in terms of the
positron energy $E_\mathrm{pos}$. We assume perfect energy
resolution and zero threshold, in order to show the shock effects
in the most favorable conditions. It is understood that such
effects will be somewhat degraded in realistic cases, depending on
detector details. Finally, just to fix the overall scale, we
assume a detector volume corresponding to 32 kton of water.
However, for the main purposes of our discussion, the event rate
units could be taken as arbitrary in the following figures.

Figure~\ref{fig07} shows our calculated positron spectra for no
oscillations ($P_{ee}=1$, dashed curve) and hypothetical full
conversion ($P_{ee}=0$, dot-dashed curve).  For partial
conversion, Eq.~(\ref{perm}) implies that the final positron
spectrum is basically a linear combination of the previous two
spectra. Examples are given in Fig.~\ref{fig07} for
$\sin^2\theta_{13}=10^{-5}$, for both normal hierarchy (dotted
curve) and inverted hierarchy (solid curve). For normal hierarchy,
the coefficients of the combination are energy-independent,
namely, $P_{ee}=\cos^2\theta_{12}$ and
$1-P_{ee}=\sin^2\theta_{12}$ [see Eq.~\ref{cases}]. Conversely,
for inverse hierarchy, such coefficients become energy- and
time-dependent through $P_H$ [$P_{ee}=\cos^2\theta_{12}P_H(E)$].
In the specific example of Fig.~\ref{fig07}, the time is fixed
through integration over a representative  bin
[$t=6\pm0.25\,\mathrm{s}$]. The energy profile of $P_H(E)$ then
governs the relative weight of the spectra $f_{\overline e}$ and
$f_{\overline x}$ in the combination. As $P_H$ decreases from its
low-energy value at the first critical energy (see
Figs.~\ref{fig03} and \ref{fig04} and related comments), the
weight of $f_{\overline x}$ increases and the positron spectrum is
shifted to higher energies. The ``shoulder'' around the peak of
the solid curve is thus an imprint of the shock passage. This
shoulder is expected to drift for time bins different from
$t=6\pm0.25\,\mathrm{s}$.

Figure~\ref{fig08} shows a set of successive positron energy
spectra, binned in time intervals of 0.5~s, from about 6 to 9
seconds after bounce. The overall decrease of the spectra for
increasing $t$ is due to the decrease in luminosity. The top panel
does not include Earth effects, while the bottom panels includes
representative Earth mantle crossing effects, calculated in the
same conditions as in \cite{Ours}. As in Fig.~\ref{fig07}, the
cases of normal and inverted hierarchy are distinguished through
dotted and solid curves, respectively, and $\sin^2\theta_{13}$ is
fixed at $10^{-5}$. The top panel shows how the shock passage can
leave an observable, time-dependent spectral deformation, in the
case of inverse hierarchy. This possibility is very exciting,
although one might realistically hope to see no more than
variations in the first spectral moments \cite{Shoc,Luna}.

From the comparison of the top and bottom panel in
Fig.~\ref{fig08}, it turns out that the spectral modulation due to
the shock might  be partly hindered, at late times, by the
additional modulation generated by Earth effects. In other words,
the ``wiggles'' produced by oscillations in the Earth might make
it difficult to identify the position of the genuine,
shock-dependent spectral deformations at different times.
Therefore, in some cases Earth effects may not necessarily
represent an additional handle \cite{Luna} to study the shock
dynamics via neutrinos.

Finally, Fig.~\ref{fig09} shows the drop in the positron event
rate (arbitrary units) as a function of time and for four
representative values of $\sin^2\theta_{13}$, ranging from
$10^{-5}$ to $10^{-2}$. Four rates are displayed in each panel,
integrated over consecutive 10 MeV energy bins. For normal
hierarchy (dotted curves), the rates decrease according to the
assumed exponential law in Eq.~(\ref{Lum}). For inverted hierarchy
(solid curves), the characteristic time structure of $P_H(t)$ (see
Fig.~\ref{fig05}) is instead reflected in a characteristic
deviation of the rate decrease with respect to a pure exponential
drop: The lower $P_H(t)$, the higher the rate (enriched in
$\overline\nu_x$ events) for  inverse hierarchy, as compared with
normal hierarchy. The two hierarchies become instead
indistinguishable when $P_H(t)\simeq 1$ (see Fig.~\ref{fig05}),
since $P_{ee}\simeq \cos^2\theta_{12}$ for both normal and
inverted neutrino mass spectra [Eq.~(\ref{cases})]. Roughly
speaking, the time structure of the event rates in
Fig.~\ref{fig09} reflects (although ``up-side-down'') the $P_H(t)$
pattern in Fig.~\ref{fig05}.

Shock-wave signatures in the time domain have been emphasized in
\cite{Schi}. It goes without saying that very high statistics
would be needed to discriminate such time structures in real
experiments.%
\footnote{A more complete and quantitative study of the
observability of shock-induced structures in supernova
(anti)neutrino signals will be performed elsewhere, in the context
of future large-volume, high-statistics detectors.}
Nevertheless, the reward could be very high. Let us consider,
e.g., the ``bathtub'' pattern of the solid curves in the lower
right panel of Fig.~\ref{fig09} ($\sin^2\theta_{13}=10^{-2}$). The
(hypothetical) experimental detection of such rather distinctive
pattern would, at the same time: ($i$) provide a snapshot (if not
a ``movie'') of the shock wave propagation; ($ii$) prove that the
neutrino mass hierarchy is inverted; and ($iii$) put a significant
lower bound on $\sin^2\theta_{13}$. Each of these results would
have a dramatic impact on our understanding of both supernova and
neutrino properties.

\section{Summary and prospects}

Supernova shock propagation can produce observable effects in the
energy and time structure of the neutrino signal, as suggested in
\cite{Schi} and also investigated in \cite{Shoc,Luna}. In the
calculation of such effects for $3\nu$ oscillations, the neutrino
crossing probability $P_H$ in the 1-3 neutrino subsystem plays an
important role. The evaluation of $P_H$ can be done either
numerically or analytically, for given shock profiles in space and
time. By assuming a simplified description of the time-dependent
density profile, we have explicitly shown that simple analytical
calculations accurately reproduce the phase-averaged numerical
results, also in the nonobvious case of multiple transitions along
the shock profile. The analytical approach has then been used to
study some relevant characteristics of the structure of $P_H$ in
energy and time. This structure can in part be reflected in
observable signals, as we have shown through some simple and
selected examples.

The simplifications used in this work, and especially the
empirical parameterization of the density profile, do not alter
our main conclusions about the validity and usefulness of a simple
analytical approach to $P_H$, as compared with brute-force
numerical calculations. However, such simplifications may affect
the specific values of $P_H$ that we have estimated at given
energy and time. These values (and the corresponding observable
effects) should not be taken too literally, since they depend on
features of the shock profile which are beyond our control. In
particular, the shape of the rarefaction zone left behind by the
shock wave should be constrained, if possible, by dedicated
supernova simulations with higher resolution in space and time.
When such issues will be clarified, more realistic calculations of
$P_H$, and of its imprint on observable neutrino signals at the
Earth, will become possible.

\acknowledgments

This work was in part supported by the Italian {\em Ministero
dell'Istruzione, Universit\`a e Ricerca\/} (MIUR) and {\em
Istituto Nazionale di Fisica Nucleare\/} (INFN) within the
``Astroparticle Physics'' research project. We thank C.~Lunardini
and T.~Janka for very useful comments. One of us (E.L.)
acknowledges kind hospitality at the Institute for Advanced Study
(Princeton, New Jersey) where this work was completed.


\begin{figure}
\vspace*{+2.4cm}\hspace*{-2.8cm}
\includegraphics[scale=0.75, bb= 30 100 500 700]{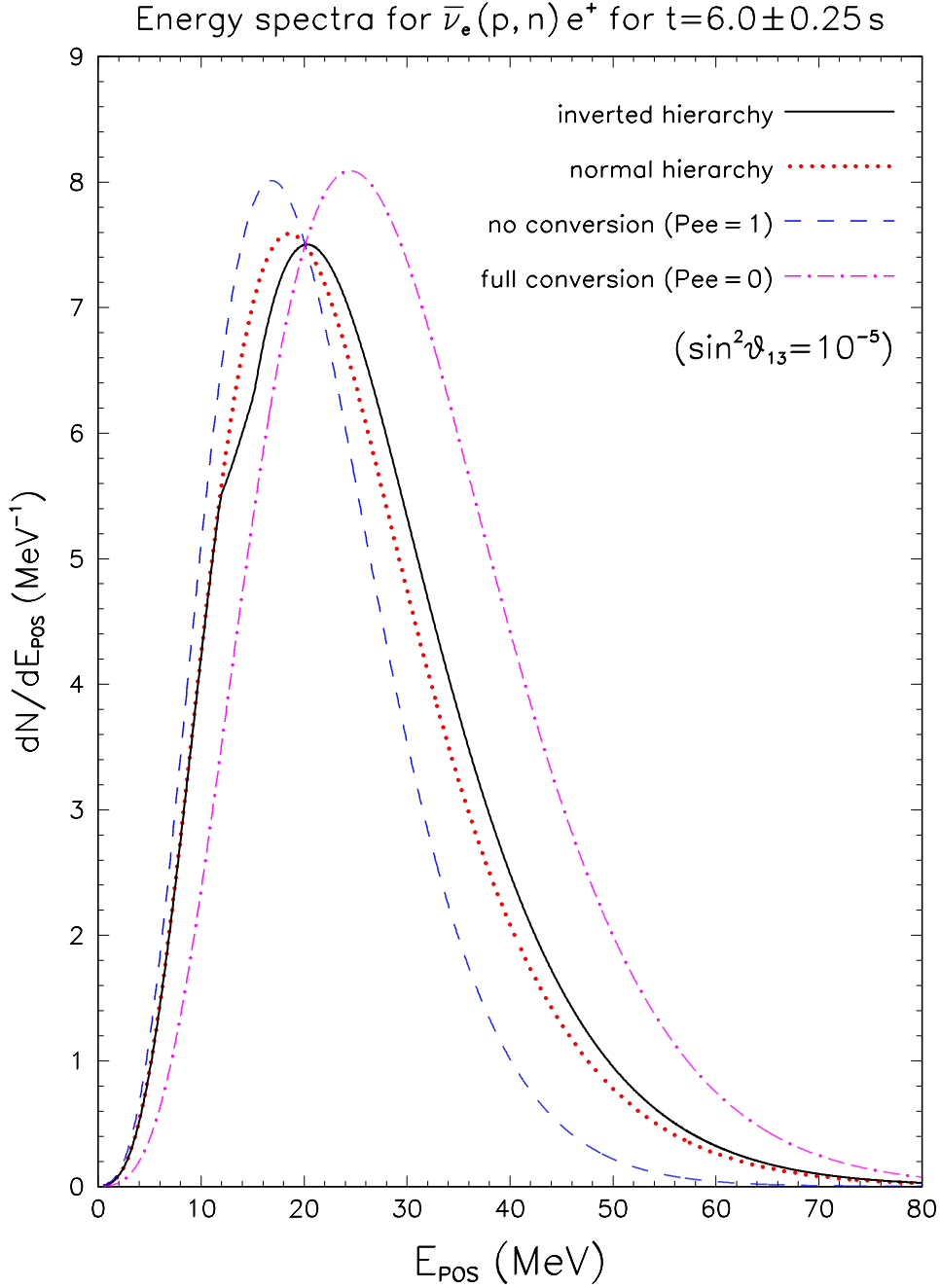}
\vspace*{2.cm} \caption{\label{fig07} Representative positron
energy spectra of supernova antineutrino events from inverse beta
decay, for different degrees of $\overline\nu_e$ conversion into
$\overline\nu_x$. Dashed curve: No conversion. Dot-dashed curve:
Full conversion. Dotted line: Partial conversion for normal
hierarchy. Solid line: Partial conversion for inverse hierarchy
(calculated for $\sin^2\theta_{13}=10^{-5}$, and integrated over
the time interval $t=6\pm 0.25\,\mathrm{s}$). The latter case
shows the shock imprint as a ``shoulder'' in the spectrum range
$E\simeq 10$--20 MeV, .}
\end{figure}
\begin{figure}
\vspace*{+2.4cm}\hspace*{-2.8cm}
\includegraphics[scale=0.75, bb= 30 100 500 700]{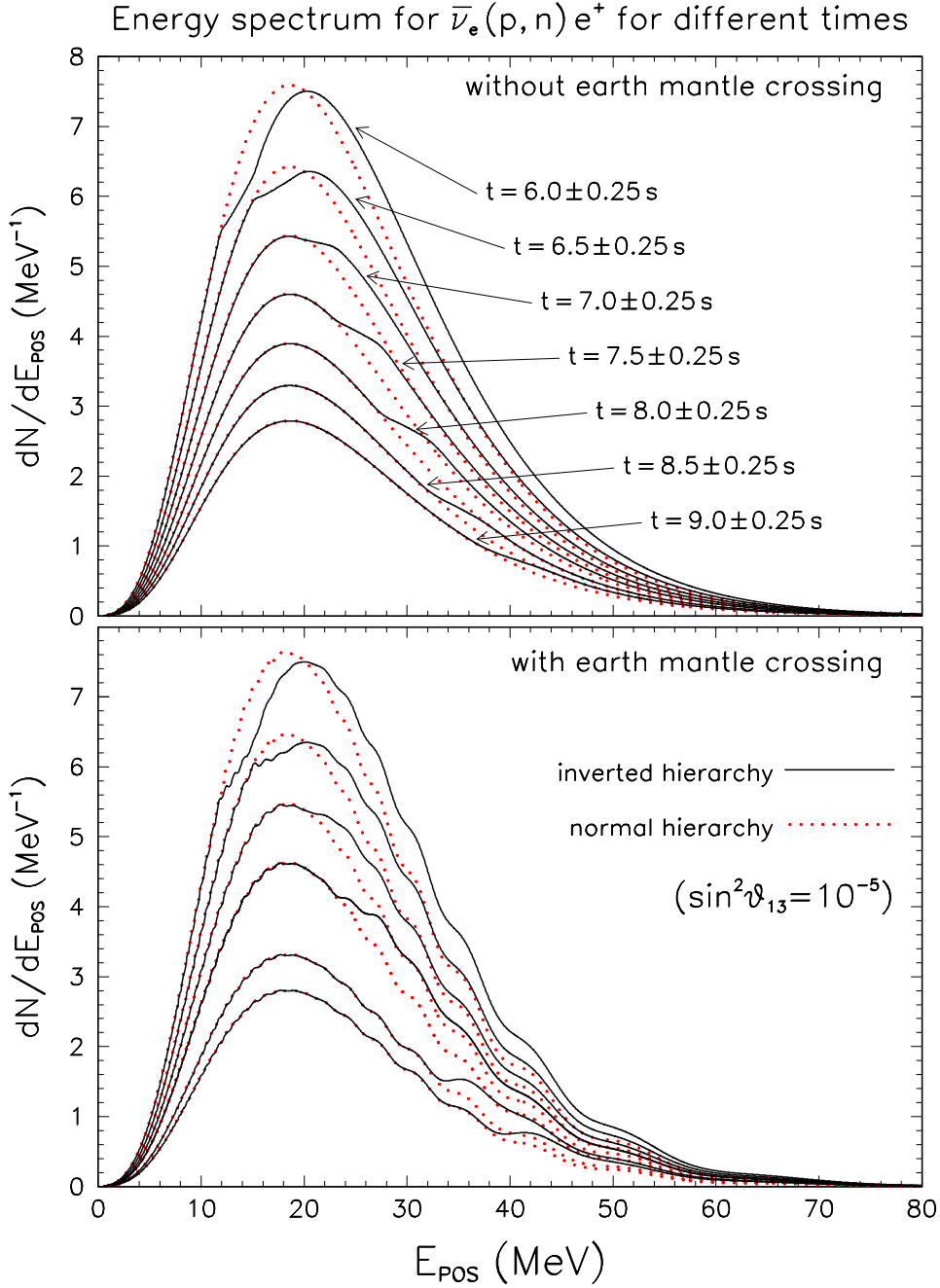}
\vspace*{2.cm} \caption{\label{fig08} Positron energy spectra for
successive time bins, calculated for $\sin^2\theta_{13}=10^{-5}$
and for normal and inverted hierarchy (dotted and solid curves,
respectively). In the case of inverted hierarchy, shock-induced
spectral deformation  move forward in energy for increasing
post-bounce time. The bottom panel includes representative effects
of Earth mantle crossing, which induce additional spectral
deformations.}
\end{figure}
\begin{figure}
\vspace*{+2.4cm}\hspace*{-2.8cm}
\includegraphics[scale=0.75, bb= 30 100 500 700]{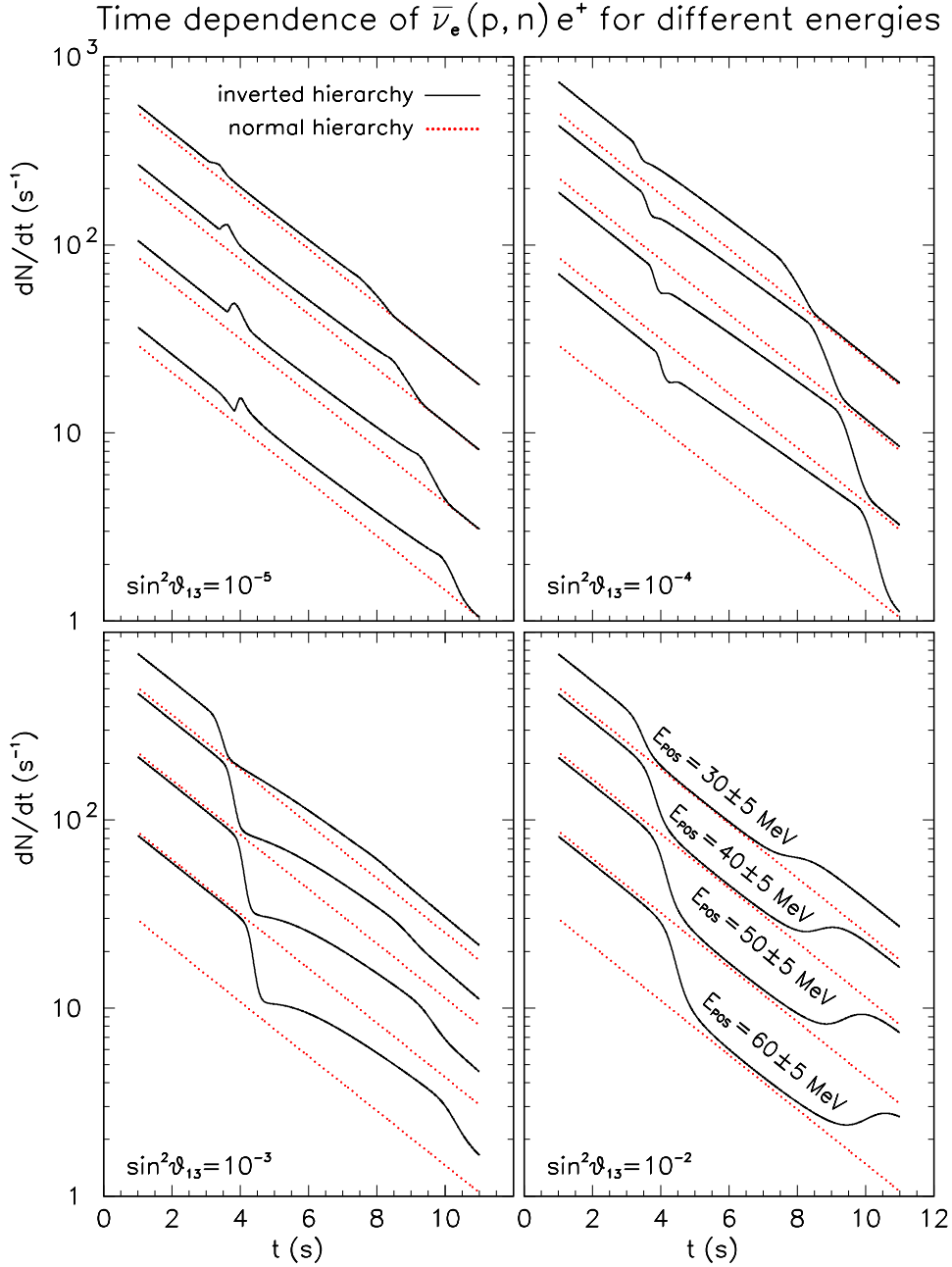}
\vspace*{2.cm} \caption{\label{fig09} Time dependence of the
positron event rate (binned in specified energy intervals) for
four representative values of $\sin^2\theta_{13}$. For normal
hierarchy (dotted lines), the event rate is just proportional to
the neutrino luminosity (assumed to decay exponentially in time).
For inverted hierarchy (solid lines) an additional time structure
is induced by the shock propagation.}
\end{figure}

\end{document}